\documentclass[twocolumn,pre,showpacs]{revtex4}
\usepackage{amssymb}
\usepackage{amsfonts}
\usepackage{amsmath}
\usepackage{graphicx}

\begin{document}

\title{Inhomogeneous diffusion and ergodicity breaking induced by global
memory effects}
\author{Adri\'{a}n A. Budini}
\affiliation{Consejo Nacional de Investigaciones Cient\'{\i}ficas y T\'{e}cnicas
(CONICET), Centro At\'{o}mico Bariloche, Avenida E. Bustillo Km 9.5, (8400)
Bariloche, Argentina, and Universidad Tecnol\'{o}gica Nacional (UTN-FRBA),
Fanny Newbery 111, (8400) Bariloche, Argentina}
\date{\today }

\begin{abstract}
We introduce a class of discrete random walk model driven by global memory
effects. At any time the right-left transitions depend on the whole previous
history of the walker, being defined by an urn-like memory mechanism. The
characteristic function is calculated in an exact way, which allows us to
demonstrate that the ensemble of realizations is ballistic. Asymptotically
each realization is equivalent to that of a biased Markovian diffusion
process with transition rates that strongly differs from one trajectory to
another. Using this \textquotedblleft inhomogeneous
diffusion\textquotedblright\ feature the ergodic properties of the dynamics
are analytically studied through the time-averaged moments. Even in the long
time regime they remain random objects. While their average over
realizations recover the corresponding ensemble averages, departure between
time and ensemble averages is explicitly shown through their probability
densities. For the density of the second time-averaged moment an ergodic
limit and the limit of infinite lag times do not commutate. All these
effects are induced by the memory effects. A generalized Einstein
fluctuation-dissipation relation is also obtained for the time-averaged
moments.
\end{abstract}

\maketitle



\section{Introduction}

Random walks dynamics are one of the more simple non-equilibrium models
which found application in diverse kind of problems arising in physics,
biology, economy, etc. In their standard Markovian formulation \cite%
{vanKampen,gardiner}, the second moment of these diffusive processes grows
linearly in time, a property shared by Brownian motion. Anomalous (sub and
super) diffusive processes \cite{bouchaud,klafterRep} depart from the
linearity condition.

The temporal dependences of the moments of a random walk are defined from an
ensemble of realizations. Nevertheless, single particle tracking microscopy
permits to define the moments from an alternative temporal moving average
performed with only one single trajectory \cite{maria,sako,iva} . From a
physical point of view, this technique allow us to ask about the ergodic
properties of a diffusion process, even when it does not have a stationary
state.

In different tracking experiments performed with biophysical arranges \cite%
{iva,unkel,simon,manzo} it was found that the diffusion coefficient (which
parametrizes the time-averaged second moment) becomes a random object that
assumes different values for each realization. This distribution of
diffusion coefficients renders the process \textit{inhomogeneous} in the
sense that in an ensemble of simple diffusers each one has a different
diffusion coefficient \cite{burov}. In addition to this feature, the
time-averaged second moments are characterized by a subdiffusive behavior.
Both properties lead to weak ergodicity breaking, that is, in contrast to
strong ergodicity breaking, time and ensemble averages differs even when the
system is able to visit the full available phase space. These striking
experimental results can be captured through a continuous-time random walk
model with waiting time distributions characterized by power-law behaviors 
\cite{burovS,burov,jeon}. These results triggered the study of the ergodic
properties of diverse anomalous diffusion process \cite%
{yan,aki,xxx,radons,cherstvy,garcia,peters,safdari,bodrova,godec,gbel,igor,fuli}
from a similar perspective.

The main goal of this paper is to explore if the inhomogeneous property of a
diffusion process (asymptotic randomness of the time-averaged moments)
jointly with its associated weak ergodicity breaking \cite{burovS,burov} may
also be induced by the presence of strong memory effects in the stochastic
dynamics. Specifically, we are interested in globally correlated dynamics,
where the walker transitions depend on its whole previous history or
trajectory.

It is known that globally correlated stochastic dynamics lead to anomalous
diffusion processes \cite%
{gunter,kim,kursten,gandi,kenkre,katja,boyer,esguerra,hanel}. On the other
hand, we remark that the interplay between memory effects and weak
ergodicity breaking was study previously such as for example in correlated
continuous-time random walk models \cite{marcin,tejedor}, single-file
diffusion \cite{file}, and fractional Brownian-Langevin motion \cite{deng}.
Here, we consider a different kind of memory processes. The model consist in
a random walker whose transitions depend on the whole previous history of
transitions. The right-left jump probabilities are defined by an urn-like
mechanism \cite{feller,norman,pitman,queen,budini}, which does not fulfill
the standard central limit theorem \cite{budini}. The ensemble dynamics
becomes superdiffusive (ballistic). Furthermore, in contrast with other
correlation mechanisms, here each realization is asymptotically equivalent
to those of a biased Markovian walker but with (random) transition rates
that assume different values for each realization. This property leads to
random time averages and its associated ergodicity breaking.

We consider a diffusive non-stationary dynamics (the statistics is not
invariant under a time shift). Similarly to the case of continuous-time
random walks (see for example Refs. \cite{burovS} and \cite{rebenStationary}%
), the studied model yields statistical laws for ergodicity breaking which
are different from those obtained from dynamics with a stationary state,
case analyzed in Ref. \cite{adrian}. In addition, here a generalized
Einstein fluctuation-dissipation relation is established \cite%
{burovS,akimoto,nogo} for the time-averaged moments.

The paper is outlined as follows. In Sec. II we introduce the stochastic
dynamics that defines the globally correlated random walk. Its ensemble
properties are studied through its characteristic function, which allows us
to calculate its moments and probability evolution. In Sec. III the
time-averaged moments and the ergodic properties are analyzed. In Sec. IV a
generalized Einstein relation is obtained from the time-averaged moments.
Section V is devoted to the Conclusions. Calculus details that support the
main results are provided in the Appendixes.

\section{Global correlated random walk dynamics}

We consider a one-dimensional random walk where both the time and position
coordinates are discrete. In each discrete time step $(t\rightarrow t+\delta
t)$ the walker perform a jump of length $\delta x$ to the right or to the
left. For simplicity, time is measured in units of $\delta t.$ Then, $%
t=0,1,2,\cdots .$ The stochastic position $X_{t}$ at time $t$ is%
\begin{equation}
X_{t}=X_{0}+\sum_{t^{\prime }=1}^{t}\sigma _{t^{\prime }}.
\end{equation}%
Here, $X_{0}$ is the initial position, and $\sigma _{t}=\pm \delta x$\ is a
random variable assigned to each step. The stochastic dynamics of the
variables $\{\sigma _{t^{\prime }}\}_{t^{\prime }=1}^{t}$ is as follows. At $%
t=1$ (first jump or transition) the two possible values are chosen with
probability%
\begin{equation}
P(\sigma _{1}=\pm \delta x)=q_{\pm },
\end{equation}%
where the weights satisfy $q_{+}+q_{-}=1.$ The next values are determinate
by a conditional probability $\mathcal{T}(\sigma _{1},\cdots \sigma
_{t}|\sigma _{t+1})$ \cite{notation} that depends on the whole previous jump
trajectory: $\sigma _{1},\cdots \sigma _{t}.$

Different memory mechanisms can be introduced through $\mathcal{T}(\sigma
_{1},\cdots \sigma _{t}|\sigma _{t+1}),$ such as for example in the elephant
random walk model \cite{gunter,kim,kursten}. Here, we analyze an alternative
urn-like dynamics \cite{adrian}, where%
\begin{equation}
\mathcal{T}(\sigma _{1},\cdots \sigma _{t}|\sigma _{t+1}=\pm \delta x)=\frac{%
\lambda q_{\pm }+t_{\pm }}{t+\lambda }.  \label{Transition}
\end{equation}%
In this expression, $\lambda $\ is a positive free dimensionless parameter.
Furthermore, $t_{+}$ and $t_{-}$ are the number of times that the walker
jumped (up to time $t$) to the right and to the left respectively, $%
t=t_{+}+t_{-}.$ Hence, with probability $\lambda /(t+\lambda )$ the walker
jumps to right or to the left with weights $q_{+}$ and $q_{-}$ respectively.
Complementarily, the jump is chosen in agreement with the weights $t_{\pm
}/(t+\lambda ),$ which gives the dependence of the dynamics over the whole
previous jump trajectory.

Notice that in the limit $\lambda \rightarrow \infty $ independent random
variables with probability $q_{\pm }$ are obtained. Hence, the stochastic
dynamics becomes an usual memoryless random walk. In the limit $\lambda
\rightarrow 0,$ the random variables $\sigma _{t}$ assume the same value as $%
\sigma _{1}.$ Therefore, a deterministic behavior follows after the first
jump.

Given the transition probability (\ref{Transition}), the set of random
variables $\{\sigma _{t}\}$ is interchangeable \cite{budini}. Therefore,
their joint probability density is invariant under arbitrary permutation of
its arguments. In consequence, the probability of the variables $\sigma _{t}$
(jump length) is independent of $t,$ $P(\sigma _{t}=\pm \delta x)=q_{\pm }.$
The average jump length reads%
\begin{equation}
\langle \sigma \rangle \equiv \int d\sigma P(\sigma )\sigma =\delta
x(q_{+}-q_{-}).  \label{Average}
\end{equation}%
Then, for $q_{+}\neq q_{-}$ a biased random walk is obtained, $\langle
\sigma \rangle \neq 0.$ The second jump moment is%
\begin{equation}
\langle \sigma ^{2}\rangle \equiv \int d\sigma P(\sigma )\sigma ^{2}=\delta
x^{2}.
\end{equation}%
Notice that both statistical moments are finite.

The initial condition $X_{0}$ jointly with the transition probability (\ref%
{Transition}) completely define the stochastic dynamics. Below, we
characterize its statistical properties.

\subsection{Characteristic function}

The stochastic process $X_{t}$\ can be described through%
\begin{equation}
x_{t}\equiv X_{t}-X_{0}=\sum_{t^{\prime }=1}^{t}\sigma _{t^{\prime }},
\end{equation}%
which measures the departure with respect to the initial condition $X_{0}.$
Its characteristic function is defined by%
\begin{equation}
Q_{t}(k)\equiv \left\langle \exp (ikx_{t})\right\rangle .  \label{Qk}
\end{equation}%
Here, $\left\langle \cdots \right\rangle $ denotes an average over an
ensemble of realizations. A close recursive relation for $Q_{t}(k)$ can be
obtained as follows. At time $t+1,$ it can be written as%
\begin{equation}
Q_{t+1}(k)=\left\langle e^{ikx_{t}}\sum_{\sigma =\pm \delta x}\mathcal{T}%
(\sigma _{1},\cdots \sigma _{t}|\sigma )e^{ik\sigma }\right\rangle .
\end{equation}%
Here, we taken into account that the random variable $\sigma _{t+1}$ is
chosen in agreement with $\mathcal{T}(\sigma _{1},\cdots \sigma _{t}|\sigma
_{t+1}).$ Notice that the average includes all possible random values of $%
\{\sigma _{i}\}_{i=1}^{i=t},$ which in turn define all possible realizations
of $x_{t}.$ From Eq.~(\ref{Transition}), we get%
\begin{eqnarray}
Q_{t+1}(k) &=&Q_{t}(k)\frac{\lambda }{t+\lambda }\sum_{\mu =\pm }q_{\mu
}e^{ik\delta x_{\mu }}  \label{previa} \\
&&+\frac{1}{t+\lambda }\sum_{\mu =\pm }\left\langle e^{ikx_{t}}t_{\mu
}\right\rangle e^{ik\delta x_{\mu }},  \notag
\end{eqnarray}%
where for shortening the expression we defined $\delta x_{\pm }\equiv \pm
\delta x.$ Given that $x_{t}=\delta x(t_{+}-t_{-}),$ the derivative of the
characteristic function (\ref{Qk}) can be written as%
\begin{equation}
\frac{d}{dk}Q_{t}(k)=i\delta x\left\langle
e^{ikx_{t}}(t_{+}-t_{-})\right\rangle .  \label{derivada}
\end{equation}%
Hence, after writing $e^{ik\delta x_{\mu }}=\cos (k\delta x_{\mu })+i\sin
(k\delta x_{\mu }),$ by using that $t=t_{+}+t_{-},$ and $q_{+}+q_{-}=1$ \cite%
{previol}, Eq. (\ref{previa})\ straightforwardly\ leads to the closed
recursive relation%
\begin{eqnarray}
Q_{t+1}(k) &=&\cos (k\delta x)Q_{t}(k)+\frac{1}{t+\lambda }\sin (k\delta x)%
\frac{1}{\delta x}\frac{d}{dk}Q_{t}(k)  \notag \\
&&+i(q_{+}-q_{-})\frac{\lambda }{t+\lambda }\sin (k\delta x)Q_{t}(k).
\label{Characteristic}
\end{eqnarray}%
This is the main result of this section. It completely characterizes the
probability and moments of $x_{t}.$

We notice that in the limit $\lambda \rightarrow 0,$ the characteristic
function is $Q_{t}(k)=\left\langle \exp (ikt\sigma _{1})\right\rangle
=q_{+}\exp (ikt\delta x)+q_{-}\exp (-ikt\delta x),$ which consistently
satisfies Eq. (\ref{Characteristic}) with $\lambda =0.$ In fact, after the
first event, the next ones assume the same value, $x_{t}=t\sigma _{1}$ [see
Eq. (\ref{Transition})]. In the limit $\lambda \rightarrow \infty ,$ the
solution of Eq. (\ref{Characteristic}) is $Q_{t}(k)=\left\langle \exp
(ik\sigma _{1})\right\rangle ^{t}=[q_{+}\exp (ik\delta x)+q_{-}\exp
(-ik\delta x)]^{t},$ which corresponds to the characteristic function of a
Markovian random walk where the steps $\sigma _{t}$ are independent random
variables.

\subsection{Moments behavior}

From the characteristic function $Q_{t}(k),$ the moments can be obtained by
differentiation as%
\begin{equation}
\langle x_{t}\rangle =-i\left. \frac{d}{dk}Q_{t}(k)\right\vert _{k=0},\ \ \
\ \ \ \ \langle x_{t}^{2}\rangle =-\left. \frac{d^{2}}{dk^{2}}%
Q_{t}(k)\right\vert _{k=0}.
\end{equation}

For the \textit{first moment,} Eq. (\ref{Characteristic}) lead to the
recursive relation%
\begin{equation}
\langle x_{t+1}\rangle =\langle x_{t}\rangle \left[ 1+\frac{1}{t+\lambda }%
\right] +\frac{\lambda }{t+\lambda }\langle \sigma \rangle ,
\end{equation}%
where the average jump length $\langle \sigma \rangle $ is given by Eq. (\ref%
{Average}). The solution of this equation is%
\begin{equation}
\langle x_{t}\rangle =\langle \sigma \rangle t=\delta x(q_{+}-q_{-})t.
\label{First}
\end{equation}%
Hence, the bias induced by $(q_{+}-q_{-})$ leads to a linear increasing of $%
\langle x_{t}\rangle .$

For the \textit{second moment,} it follows the recursive relation%
\begin{equation}
\langle x_{t+1}^{2}\rangle =\langle x_{t}^{2}\rangle \left[ 1+\frac{2}{%
t+\lambda }\right] +\frac{2\lambda }{t+\lambda }\langle x_{t}\rangle \langle
\sigma \rangle +\langle \sigma ^{2}\rangle ,
\end{equation}%
whose solution is given by%
\begin{equation}
\langle x_{t}^{2}\rangle =\frac{\langle \sigma ^{2}\rangle }{1+\lambda }%
(t^{2}+t\lambda )+\frac{\langle \sigma \rangle ^{2}\lambda }{1+\lambda }%
(t^{2}-t).  \label{Second}
\end{equation}

From Eqs. (\ref{First}) and (\ref{Second}), the second centered moment reads%
\begin{equation}
\langle x_{t}^{2}\rangle -\langle x_{t}\rangle ^{2}=\left[ \frac{\langle
\sigma ^{2}\rangle -\langle \sigma \rangle ^{2}}{1+\lambda }\right]
(t^{2}+t\lambda ).  \label{third}
\end{equation}%
Hence, the memory effects leads to a \textit{superdiffusive} behavior, which
in the asymptotic time regime becomes ballistic. The ballistic regime is
valid at any time when $\lambda \rightarrow 0.$ Consistently, in the limit $%
\lambda \rightarrow \infty $ (memoryless case) it follows%
\begin{equation}
\langle x_{t}^{2}\rangle -\langle x_{t}\rangle ^{2}=[\langle \sigma
^{2}\rangle -\langle \sigma \rangle ^{2}]t,  \label{normal}
\end{equation}%
which corresponds to an expected standard diffusive behavior.

\subsection{Probability evolution}

After Fourier inversion, the characteristic function leads to a recursive
relation for the probability $P_{t}(x)$ of $x_{t}.$ We get \cite{fourier}%
\begin{equation}
P_{t+1}(x)=W_{t}^{+}P_{t}(x-\delta x)+W_{t}^{-}P_{t}(x+\delta x),
\label{masterEq}
\end{equation}%
where%
\begin{equation}
W_{t}^{\pm }=\frac{1}{2}\left\{ 1\pm \frac{1}{t+\lambda }\left[ (\frac{x\mp
\delta x}{\delta x})+\lambda (q_{+}-q_{-})\right] \right\} .
\end{equation}%
The evolution (\ref{masterEq}), which is valid for $t\geq 1,$ describes a
hopping process with transitions $W_{t}^{\pm }.$ In the limit $\lambda
\rightarrow \infty ,$ it follows $W_{t}^{\pm }=q_{\pm },$ recovering a
standard random walk. For finite $\lambda ,$ the memory effects appears
through $W_{t}^{\pm }.$ Furthermore, for $X_{t}$ the hopping also depends on
the initial condition $(x\rightarrow X-X_{0}),$ non-Markovian property
shared by the elephant random walk model \cite{gunter}.

An interesting aspect of the evolution (\ref{masterEq}) is given by its 
\textit{continuous limit.} It follows by taking the limits in which both the
length jump $(\delta x\rightarrow 0)$ and the time interval between jumps $%
(\delta t\rightarrow 0)$ vanish. Then, we can approximate (for simplicity
the (dimensional) continuous time is also denoted by $t$)%
\begin{equation}
P_{t}(x\mp \delta x)\rightarrow P_{t}(x)\mp \delta x\frac{\partial }{%
\partial x}P_{t}(x)+\frac{\delta x^{2}}{2}\frac{\partial ^{2}}{\partial x^{2}%
}P_{t}(x),
\end{equation}%
jointly with%
\begin{equation}
P_{t+1}(x)-P_{t}(x)\rightarrow \delta t\frac{\partial }{\partial t}P_{t}(x).
\end{equation}%
Introducing these approximations in Eq. (\ref{masterEq}), it follows the
equation%
\begin{eqnarray}
\frac{\partial }{\partial t}P_{t}(x) &=&D\frac{\partial ^{2}}{\partial ^{2}x}%
P_{t}(x)-\frac{1}{t+t_{\lambda }}\frac{\partial }{\partial x}[xP_{t}(x)] 
\notag \\
&&-\frac{t_{\lambda }}{t+t_{\lambda }}V\frac{\partial }{\partial x}P_{t}(x),
\label{FP}
\end{eqnarray}%
where the parameters are%
\begin{equation}
D\equiv \frac{1}{2}\frac{\delta x^{2}}{\delta t},\ \ \ \ \ \ \ V\equiv
(q_{+}-q_{-})\frac{\delta x}{\delta t},\ \ \ \ \ \ \ t_{\lambda }\equiv
\lambda \delta t.  \label{Continuo}
\end{equation}

The Fokker-Planck equation (\ref{FP}) corresponds to a Brownian particle
driven by a harmonic potential with spring constant $1/(t+t_{\lambda }).$ A
similar result was obtained in Ref. \cite{gunter}\ for the elephant random
walk model.

In the limit $\lambda \rightarrow \infty ,$ Eq. (\ref{FP}) becomes%
\begin{equation}
\frac{\partial }{\partial t}P_{t}(x)=D\frac{\partial ^{2}}{\partial ^{2}x}%
P_{t}(x)-V\frac{\partial }{\partial x}P_{t}(x).
\end{equation}%
Consistently, this equation corresponds to the probability evolution of a
Brownian particle with diffusion coefficient $D$ and subjected to a constant
force proportional to $V.$

The evolution Eq. (\ref{FP}) also leads to a superdiffusive ballistic
process. Its solution can be written as $[P_{t=0}(x)=\delta (x)]$%
\begin{equation}
P_{t}(x)=\sqrt{\frac{1}{2\pi \sigma _{t}^{2}}}\exp \left[ -\frac{(x-Vt)^{2}}{%
2\sigma _{t}^{2}}\right] ,\ \ \ \ \ \ \sigma _{t}^{2}\equiv 2\frac{D}{%
t_{\lambda }}t(t+t_{\lambda }).
\end{equation}%
Hence, the (time dependent) harmonic potential is unable to induce a (time
independent) stationary state. In the limit $\lambda \rightarrow 0,$ the
previous solution reads $P_{t}(x)=\delta (x-Vt).$

\section{Inhomogeneous diffusion and Ergodicity breaking}

The ergodic properties of a time series $X(t)$ associated to an arbitrary
random walker can be analyzed through the \textit{time-averaged moments} 
\cite{burovS,burov}, which are definedby the following temporal moving
average%
\begin{equation}
\delta _{\kappa }(t,\Delta )\equiv \frac{\int_{0}^{t-\Delta }dt^{\prime
}[X(t^{\prime }+\Delta )-X(t^{\prime })]^{\kappa }}{t-\Delta }.
\label{TimeAverage}
\end{equation}%
Here, $\Delta $ is called the lag (or delay) time, and $\kappa $ is a
natural number, $\kappa =1,2,\cdots .$

For \textit{ergodic diffusion processes,} in the limit of increasing times, $%
\delta _{\kappa }(t,\Delta )$ recovers the ensemble behavior of the
corresponding moments, that is 
\begin{equation}
\delta _{\kappa }(\Delta )\equiv \lim {}_{t\rightarrow \infty }\delta
_{\kappa }(t,\Delta )=\langle \lbrack X(\Delta )-X(0)]^{\kappa }\rangle .
\label{Ergodicity}
\end{equation}%
Here, the initial condition $X(0)$ follows from the translational invariance
of Eq. (\ref{TimeAverage}). A weaker condition can be formulated by
demanding the equality of the asymptotic behaviors $(\Delta \rightarrow
\infty )$ of both terms in Eq. (\ref{Ergodicity}).

Non-ergodic process do not fulfill Eq. (\ref{Ergodicity}). In particular,
inhomogeneous diffusion corresponds to the case in which $\delta _{\kappa
}(t,\Delta ),$ even in the long time limit, becomes a random object that
assumes different values for each particular realization. Below we study the
time-averaged moments $\delta _{\kappa }(t,\Delta )$ for the random walk
introduced in the previous section.

\subsection{Asymptotic randomness}

For the proposed model, given that the permanence time in each state is
finite, a central ingredient that determines its ergodic properties is the
asymptotic behavior $(\lim_{t\rightarrow \infty })$ of the transition
probability $\mathcal{T}(\sigma _{1},\cdots \sigma _{t}|\sigma _{t+1}=\pm
\delta x).$ For the urn model, Eq. (\ref{Transition}), it is known that it
converges\ to random values $f_{\pm }$ \cite{queen,adrian}, that is,%
\begin{equation}
\lim_{t\rightarrow \infty }\mathcal{T}(\sigma _{1},\cdots \sigma _{t}|\sigma
_{t+1}=\pm \delta x)=f_{\pm },  \label{fractions}
\end{equation}%
where $0\leq f_{\pm }\leq 1$ and $f_{+}+f_{-}=1.$ In each particular
realization $f_{\pm }$ assume different random values. Their probability
density $\mathcal{P}(f_{\pm })$ is a Beta distribution \cite{queen,adrian}%
\begin{equation}
\mathcal{P}(f_{\pm })=\frac{\Gamma (\lambda )}{\Gamma (\lambda _{+})\Gamma
(\lambda _{-})}f_{+}^{\lambda _{+}-1}f_{-}^{\lambda _{-}-1},  \label{Betal}
\end{equation}%
where $\lambda _{\pm }\equiv \lambda q_{\pm },$ and $\Gamma (x)$ is the
Gamma function. For clarity, these results are rederived in Appendix A. The
average over realizations of $f_{\pm }$ is $\left\langle f_{\pm
}\right\rangle =\int_{0}^{1}df_{+}\ \mathcal{P}(f_{\pm })f_{\pm }=q_{\pm }.$
For alternative memory mechanisms, such as that associated to the elephant
random walk model \cite{gunter,kim,kursten}, the previous randomness is
absent \cite{adrian}.

The convergence of the transition probability to random values
straightforwardly lead to an inhomogeneous diffusion process. In fact, each
realization becomes equivalent to that of a biased Markovian random walk
process with transition rates $f_{\pm }.$ The bias arises because (even when 
$q_{+}=q_{-})$ in general $f_{+}\neq f_{-}.$

In the limit $\lambda \rightarrow \infty ,$ from Eq. (\ref{Betal}) it
follows $\mathcal{P}(f_{\pm })=\delta (f_{\pm }-q_{\pm }),$ implying that
the fractions $f_{\pm },$ at any stage of the diffusion process, assume
deterministically the values $q_{\pm }.$ This case corresponds to the
absence of memory and leads to a standard diffusion process [defined by Eq. (%
\ref{normal})].

The asymptotic property (\ref{fractions}) implies that at large times the
time-averaged moment $\delta _{\kappa }(\Delta )=\lim {}_{t\rightarrow
\infty }\delta _{\kappa }(t,\Delta )$ becomes a random variable. In fact,
its average over realizations can be written as%
\begin{equation}
\left\langle \delta _{\kappa }(\Delta )\right\rangle =\left\langle \delta
_{\kappa }(\Delta ,f_{\pm })\right\rangle =\int_{0}^{1}df_{+}\ \mathcal{P}%
(f_{\pm })\delta _{\kappa }(\Delta ,f_{\pm }).
\end{equation}%
In this expression $\delta _{\kappa }(\Delta ,f_{\pm })$ corresponds to the
(asymptotic) time-averaged moment corresponding to a memoryless random walk
with transition rate $\mathcal{T}(\sigma _{1},\cdots \sigma _{t}|\sigma
_{t+1}=\pm \delta x)=f_{\pm }.$ Given the ergodicity of this kind of
dynamics, under the replacements $q_{\pm }\rightarrow f_{\pm },$ $%
t\rightarrow \Delta ,$ from Eqs. (\ref{Ergodicity}) and Eq. (\ref{First}) we
get%
\begin{equation}
\delta _{1}(\Delta ,f_{\pm })=\Delta \delta x(f_{+}-f_{-}).  \label{Delta1}
\end{equation}%
Similarly, taking the limit $\lambda \rightarrow \infty $ (memoryless case)
and under the same replacements, from Eq. (\ref{Second}) we get%
\begin{equation}
\delta _{2}(\Delta ,f_{\pm })=\delta x^{2}\{(f_{+}-f_{-})^{2}\Delta
^{2}+[1-(f_{+}-f_{-})^{2}]\Delta \}.  \label{Delta2}
\end{equation}

Eqs. (\ref{Delta1}) and (\ref{Delta2}) define the random values (written in
terms of $f_{\pm }$) that assume the time-averaged moments (\ref{TimeAverage}%
) in the long time limit. In order to check these results, in Fig. 1 we plot 
$\delta _{1}(t,\Delta )$ for the global correlated random walk defined by
Eq. (\ref{Transition}). From each generated realization, $\delta
_{1}(t,\Delta )$ is obtained from its definition Eq. (\ref{TimeAverage}).
Consistently with the analysis, each curve (for $\Delta <t$) can be very
well fitted by the approximation (\ref{Delta1}), that is, a linear behavior
in $\Delta $ is observed.

In Fig. 2, for a unbiased random walk $(q_{1}=q_{2}),$ we plot different
realizations corresponding to the second time-averaged moment $\delta
_{2}(t,\Delta ).$ Consistently with Eq. (\ref{Delta2}) a quadratic behavior
is observed for $\Delta <t.$%
\begin{figure}[tbp]
\includegraphics[bb=57 286 417 560,angle=0,width=8.5cm]{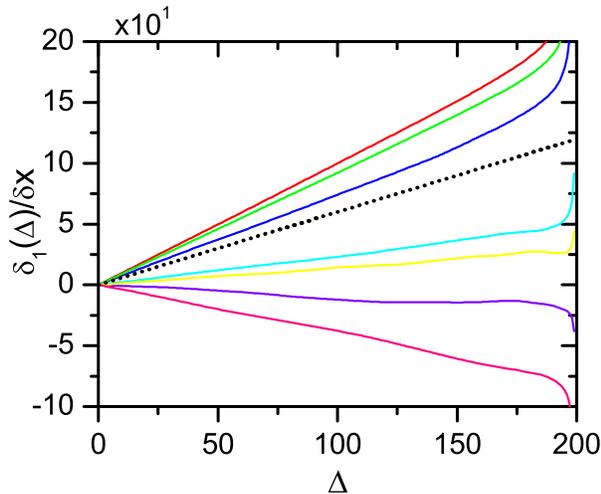}
\caption{Different realizations (full lines) of the first time-averaged
moment $\protect\delta _{1}(t,\Delta )$ [Eq. (\protect\ref{TimeAverage})]
corresponding to the globally correlated random walk dynamics defined by Eq.
(\protect\ref{Transition}). The parameters are $\protect\lambda =2,$ $%
q_{+}=0.8,$ $q_{-}=0.2,$ and $t=200.$ The dotted (black) line corresponds to
the analytical expression (\protect\ref{Delta1Media}), which gives the
ensemble mean value.}
\end{figure}

For both $\delta _{1}(t,\Delta )$ and $\delta _{2}(t,\Delta )$ the behaviors
predicted by Eqs. (\ref{Delta1}) and (\ref{Delta2}) loss their validity when 
$\Delta \approx t.$ In fact, in both figures an appreciable deviation can be
observed in that regime. The fraction of (lag) time $\Delta $ over which
that happens diminishes for increasing $t.$

\subsection{Ergodicity in mean value}

The previous figures explicitly show that, contrarily to ergodic dynamics,
here the memory effects lead to a randomness of the time-averaged moments.
Their average over an ensemble of realizations can be performed by using the
probability distribution (\ref{Betal}). Using that $\left\langle f_{\pm
}\right\rangle =\int_{0}^{1}df_{+}\ \mathcal{P}(f_{\pm })f_{\pm }=q_{\pm },$
Eq. (\ref{Delta1}) leads to%
\begin{equation}
\left\langle \delta _{1}(\Delta )\right\rangle =\Delta \langle \sigma
\rangle .  \label{Delta1Media}
\end{equation}%
Furthermore, using that $\left\langle (f_{+}-f_{-})^{2}\right\rangle
=[1+\lambda (q_{+}-q_{-})^{2}]/(1+\lambda ),$ from Eq. (\ref{Delta2}) it
follows%
\begin{equation}
\left\langle \delta _{2}(\Delta )\right\rangle =\frac{\langle \sigma
^{2}\rangle }{1+\lambda }(\Delta ^{2}+\Delta \lambda )+\frac{\langle \sigma
\rangle ^{2}\lambda }{1+\lambda }(\Delta ^{2}-\Delta ).  \label{Delta2Media}
\end{equation}%
The last two expressions, under the replacement $\Delta \rightarrow t$
recover Eqs. (\ref{First}) and (\ref{Second}) respectively. Thus, the first
two moments satisfy the ergodicity condition (\ref{Ergodicity}) only when
averaged over realizations 
\begin{figure}[t]
\includegraphics[bb=57 286 417 560,angle=0,width=8.5cm]{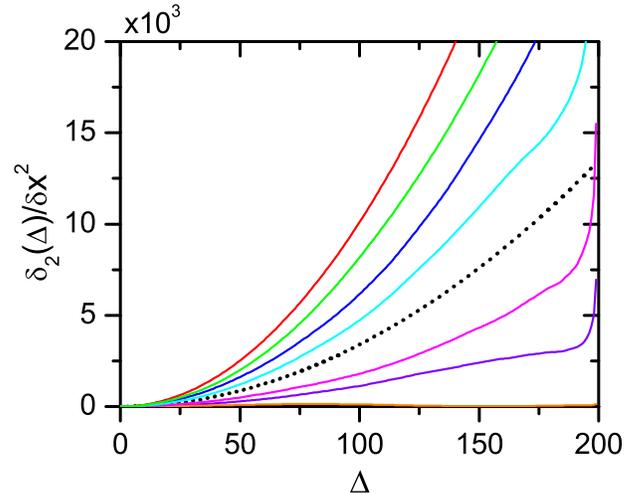} 
\caption{Different realizations (full lines) of the second time-averaged
moment $\protect\delta _{2}(t,\Delta )$ [Eq. (\protect\ref{TimeAverage})]
corresponding to an unbiased globally correlated random walk dynamics. The
parameters are $\protect\lambda =2,$ $q_{+}=q_{-}=1/2,$ and $t=200.$ The
dotted (black) line corresponds to the analytical expression (\protect\ref%
{Delta2Media}), which gives their ensemble mean value.}
\end{figure}
The validity of both results, Eqs. (\ref{Delta1Media}) and (\ref{Delta2Media}%
), was checked numerically. In Fig. 3, the solid black lines are defined by
these equations, while the circles correspond to an average over
realizations, such as those shown in Figs. (1) and (2).

Interestingly, the previous property is also valid for higher time-averaged
moments,%
\begin{equation}
\langle \delta _{\kappa }(\Delta )\rangle =\lim {}_{t\rightarrow \infty
}\langle \delta _{\kappa }(t,\Delta )\rangle =\langle \lbrack
X(t)-X(0)]^{\kappa }\rangle |_{t=\Delta }.  \label{MeanErgodic}
\end{equation}%
Thus, in terms of the characteristic function (\ref{Qk}) they can be written
as%
\begin{equation}
\left\langle \delta _{\kappa }(\Delta )\right\rangle =i^{-\kappa }\left. 
\frac{d^{\kappa }}{dk^{\kappa }}Q_{\Delta }(k)\right\vert _{k=0}.
\label{MeanErgodicBis}
\end{equation}%
The equality (\ref{MeanErgodic}) is demonstrated in Appendix B. We notice
that for an arbitrary stochastic signal $X(t)$\ we may consider the equality
(\ref{MeanErgodic}) as a definition of ergodicity in mean value.

\subsection{Probability densities}

While the asymptotic value $\delta _{\kappa }(\Delta )=\lim {}_{t\rightarrow
\infty }\delta _{\kappa }(t,\Delta )$ of the time-averaged moments is
random, Eq. (\ref{MeanErgodic}) say us that their average over realizations
recover the ensemble behavior. Therefore, we can affirm that the random
walker is ergodic in average. The lack of ergodicity is given by the random
nature of\ $\delta _{\kappa }(\Delta ).$ In fact, higher moments $%
\left\langle [\delta _{\kappa }(\Delta )]^{n}\right\rangle $ $(n\geq 2)$ can
not be related with the ensemble behavior. In order to characterize the lack
of ergodicity, we introduce the \textit{normalized (asymptotic)
time-averaged moments}%
\begin{equation}
\xi _{\kappa }\equiv \lim_{t\rightarrow \infty }\frac{\delta _{\kappa
}(t,\Delta )}{\left\langle \delta _{\kappa }(t,\Delta )\right\rangle }=\frac{%
\delta _{\kappa }(\Delta ,f_{\pm })}{\left\langle \delta _{\kappa }(\Delta
)\right\rangle },  \label{normalizedMoments}
\end{equation}%
their probability density being denoted by $P(\xi _{\kappa }).$ Ergodicity
in probability density corresponds to the absence of randomness,%
\begin{equation}
P(\xi _{\kappa })=\delta (\xi _{\kappa }-1).  \label{PErgodic}
\end{equation}%
%
%
%
%
%
%
%
%
%
%
%
%
%
%
%
%
%
%
%
%
%
%
%
%
%
%
%
%
%
%
%
%
%
%
%
%
%
%
%
%
%
%
\begin{figure}[t]
\includegraphics[bb=24 377 552 604,angle=0,width=8.5cm]{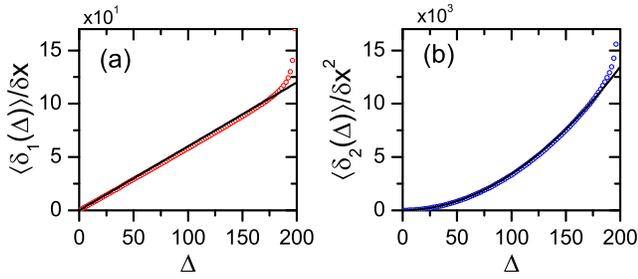}
\caption{Average over realizations of the first (a) and second (b)
time-averaged moments $\protect\delta _{1}(t,\Delta )$ and $\protect\delta %
_{2}(t,\Delta ).$ The parameters are the same than in Figs. 1 and 2
respectively. The circles correspond to a numerical average performed with $%
10^{3}$ realizations. The full lines correspond to Eqs. (\protect\ref%
{Delta1Media}) and (\protect\ref{Delta2Media}) respectively.}
\end{figure}

For $\kappa =1,$\ from Eqs. (\ref{Delta1}) and (\ref{Delta1Media}) we get%
\begin{equation}
\xi _{1}=\frac{(f_{+}-f_{-})}{(q_{+}-q_{-})},  \label{EpsiOne}
\end{equation}%
which is a random variable independent of $\Delta .$ It characterizes the
asymptotic (random) bias of the globally correlated random walk. Its
probability distribution, from Eq. (\ref{Betal}) reads%
\begin{equation}
P(\xi _{1})=\frac{1}{\mathcal{N}}|\delta q|(1+\delta q\xi _{1})^{\lambda
_{+}-1}(1-\delta q\xi _{1})^{\lambda _{-}-1}.  \label{DensityOne}
\end{equation}%
Here, $\delta q\equiv q_{+}-q_{-},$ and as before $\lambda _{\pm }=\lambda
q_{\pm }.$ The normalization constant is $\mathcal{N}=2^{\lambda -1}\Gamma
(\lambda _{+})\Gamma (\lambda _{-})/\Gamma (\lambda ).$ The density has
support in the interval defined by $|\xi _{1}|\leq 1/|\delta q|,$ and
consistently with the definition (\ref{normalizedMoments}) satisfies $%
\langle \xi _{1}\rangle =\int_{-1/|\delta q|}^{+1/|\delta q|}P(\xi _{1})\xi
_{1}d\xi _{1}=1.$ Furthermore, for $\lambda <\infty $ it departs from Eq. (%
\ref{PErgodic}).

In Fig. 4 we plot a set of probability densities $P(\xi _{1})$ jointly with
their numerical versions. They were determinate from a set of realizations
such as those shown in Fig. 1. The analytical expressions fit very well the
numerical results. Depending on the memory parameter $\lambda ,$\ the
density develops very different dependences. For increasing $\lambda ,$ the
density is peaked around one [see Fig. 4(d)], which indicates that the
ergodic regime is approached.%
\begin{figure}[tbp]
\includegraphics[bb=43 590 713 1140,angle=0,width=8.5cm]{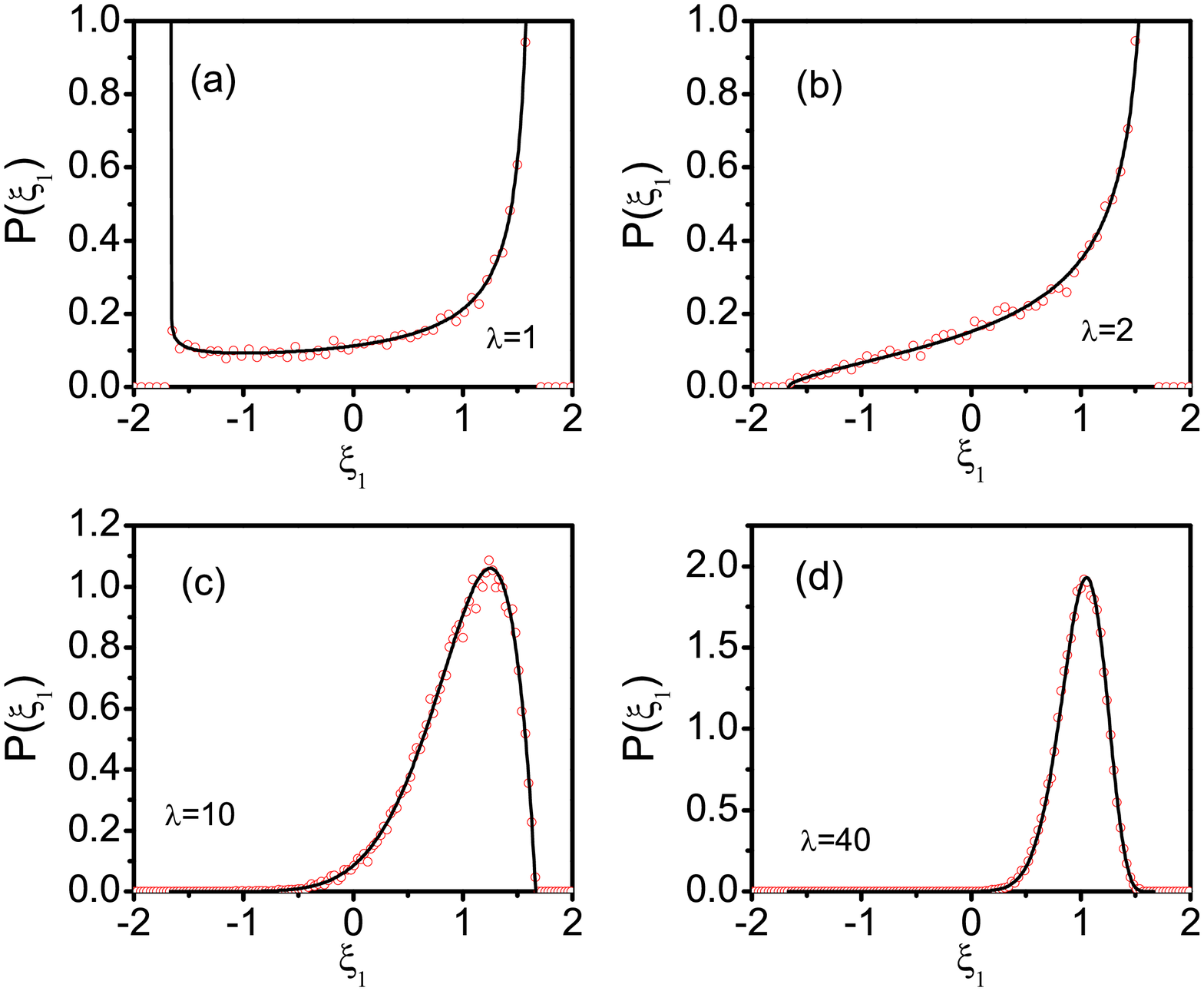}
\caption{Probability density $P(\protect\xi _{1})$ corresponding to the
normalized first time-averaged moment, Eq. (\protect\ref{normalizedMoments})
with $\protect\kappa =1.$ The full lines correspond to the analytical result
Eq. (\protect\ref{DensityOne}). The circles correspond to a numerical
simulations with $10^{4}$ realizations. The parameters are $q_{+}=0.8,$ $%
q_{-}=0.2,$ $\Delta =100,$ and $t=1000.$ In (a) $\protect\lambda =1,$ (b) $%
\protect\lambda =2,$ (c) $\protect\lambda =10,$ and in (d) $\protect\lambda %
=40.$}
\end{figure}

The second normalized moment $[\kappa =2$ in Eq. (\ref{normalizedMoments}%
)],\ from Eq. (\ref{Delta2}) can be written as%
\begin{equation}
\xi _{2}=a(f_{+}-f_{-})^{2}+b,  \label{IpsilonTwo}
\end{equation}%
where $a$ and $b$ are functions that also follows from Eq. (\ref{Delta2})
and only depend on $\Delta \ $and\ $\lambda .$ From Eq. (\ref{Betal}) we get
the probability density%
\begin{equation}
P(\xi _{2})=\frac{1}{\mathcal{N}}\frac{1}{|a|}\sqrt{\frac{a}{\xi _{2}-b}}%
\left( 1-\frac{\xi _{2}-b}{a}\right) ^{\frac{\lambda }{2}-1}.
\label{DensityTwo}
\end{equation}%
The variable $\xi _{2}$ take values in the interval $(b,a+b).$ Consistently
with Eq. (\ref{normalizedMoments}), it satisfies $\langle \xi _{2}\rangle
=\int_{b}^{a+b}P(\xi _{2})\xi _{2}d\xi _{2}=b+a/(1+\lambda )=1.$

For an unbiased random walk, $q_{+}=q_{-}=1/2,$ we obtain $\mathcal{N}%
=2^{\lambda -1}\Gamma ^{2}(\lambda /2)/[\Gamma (\lambda )],$ while from Eq. (%
\ref{Delta2Media}) it follows%
\begin{equation}
a=\frac{(\Delta -1)(1+\lambda )}{\Delta +\lambda },\ \ \ \ \ \ \ \ \ \ \ \ \
b=\frac{1+\lambda }{\Delta +\lambda },
\end{equation}%
which satisfy the previous condition $b+a/(1+\lambda )=1.$%
%
%
%
\begin{figure}[tbp]
\includegraphics[bb=43 590 713 1140,angle=0,width=8.5cm]{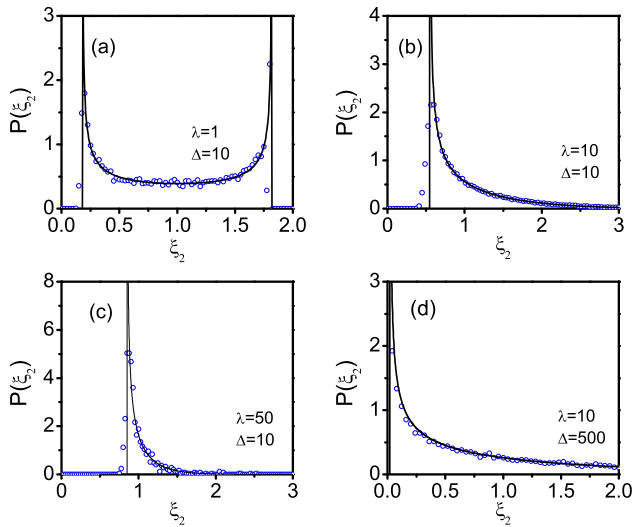}
\caption{Probability density $P(\protect\xi _{2})$ corresponding to the
normalized second time-averaged moment, Eq. (\protect\ref{normalizedMoments}%
) with $\protect\kappa =2.$ The full lines correspond to the analytical
result Eq. (\protect\ref{DensityTwo}). The circles correspond to a numerical
simulations with $5\times 10^{4}$ realizations. The parameters are $%
q_{+}=q_{-}=1/2,$ and $t=1000.$ In (a) $\protect\lambda =1,$ $\Delta =10,$
(b) $\protect\lambda =2,$ $\Delta =10,$ (c) $\protect\lambda =10,$ $\Delta
=10,$ and in (d) $\protect\lambda =40,$ $\Delta =500.$}
\end{figure}

In the limit $\lambda \rightarrow \infty $ (with finite $\Delta ),$ the
density $P(\xi _{2})$ becomes a delta Dirac function%
\begin{equation}
\lim_{\lambda \rightarrow \infty }P(\xi _{2})=\delta (\xi _{2}-1),
\label{DeltaErgo2}
\end{equation}%
which corresponds to the ergodic regime. This results follow
straightforwardly from Eqs. (\ref{IpsilonTwo}) and (\ref{Betal}). On the
other hand, in the limit $\Delta \rightarrow \infty $ (with finite $\lambda
),$ the parameter $a$ goes to $1+\lambda ,$ while $b$ vanishes. Hence,%
\begin{equation}
\lim_{\Delta \rightarrow \infty }P(\xi _{2})=\frac{1}{\mathcal{N}}\sqrt{%
\frac{1}{(1+\lambda )\xi _{2}}}\left[ 1-\frac{\xi _{2}}{1+\lambda }\right] ^{%
\frac{\lambda }{2}-1}.  \label{DelayInfinito}
\end{equation}%
From here, it is simple to proof that both kind of limits do not commutate,%
\begin{equation}
\lim_{\Delta \rightarrow \infty }\lim_{\lambda \rightarrow \infty }P(\xi
_{2})\neq \lim_{\lambda \rightarrow \infty }\lim_{\Delta \rightarrow \infty
}P(\xi _{2}).  \label{NoConmutan}
\end{equation}%
In fact,%
\begin{equation}
\lim_{\Delta \rightarrow \infty }\lim_{\lambda \rightarrow \infty }P(\xi
_{2})=\delta (\xi _{2}-1),
\end{equation}%
while from Eq. (\ref{DelayInfinito}) we get the Gamma density%
\begin{equation}
\lim_{\lambda \rightarrow \infty }\lim_{\Delta \rightarrow \infty }P(\xi
_{2})=\sqrt{\frac{1}{2\pi \xi _{2}}}\exp \left[ -\frac{\xi _{2}}{2}\right] .
\label{GamaNegativas}
\end{equation}%
In spite of this difference, notice that the previous two probability
densities lead to $\langle \xi _{2}\rangle =1.$

In\ order to check the previous results, in Fig. 5 we plot $P(\xi _{2})$
obtained numerically from a set of realizations such as those shown in Fig.
2. For $\lambda \lesssim 1,$ the distribution assume a $U$-like form [Fig.
5(a)]. For higher values of $\lambda ,$ added to the power-law behavior
predicted by Eq. (\ref{DensityTwo}) [Fig. 5(b)], $P(\xi _{2})$ approaches a
delta Dirac function [Fig. 5(c)] centered in $\xi _{2}=1,$ Eq. (\ref%
{DeltaErgo2}). When $\Delta \gg \lambda ,$ the distribution approaches the
limit defined by Eq. (\ref{DelayInfinito}), Fig. 5(d), which in the scale of
the plot is almost indistinguishable from the behavior (\ref{GamaNegativas}%
). Therefore, Fig. 5 (c) and 5(d) explicitly show the fact that in general
the ergodic limit and the limit of infinite delay times do not commutate for
the normalized moments.

\subsection{Correlations between time-averaged moments}

In the previous section we characterized the probabilities densities of the
asymptotic first and second time-averaged moments. It is interesting to note
that these objects are correlated between them. In fact, from Eqs. (\ref%
{Delta1}) and (\ref{Delta2}) it is possible to obtain the relation $\delta
_{2}(\Delta ,f_{\pm })=[\delta _{1}(\Delta ,f_{\pm })]^{2}(1-1/\Delta
)+\delta x^{2}\Delta ,$ which implies that%
\begin{equation}
\lim_{t\rightarrow \infty }\delta _{2}(t,\Delta )=\lim_{t\rightarrow \infty
}[\delta _{1}(t,\Delta )]^{2}\left( 1-\frac{1}{\Delta }\right) +\delta
x^{2}\Delta .  \label{correlacionadas}
\end{equation}%
Therefore, in the long time limit, the realizations of $\delta _{1}(t,\Delta
)$ and $\delta _{2}(t,\Delta )$\ becomes proportional. The realizations
shown in Figs. (1) and (2) are consistent with this relation, which is
strictly valid in the limit $t\rightarrow \infty .$ In spite of this fact,
due to their different scaling with $\Delta ,$ in the long time regime their
probabilities densities develop very different behaviors [see Eqs. (\ref%
{DensityOne}) and (\ref{NoConmutan})]. Relations like that defined by Eq. (%
\ref{correlacionadas}) also appear in higher time-averaged moments. In fact,
for all of them, their asymptotic behavior can always be written in terms of
the random variables $f_{\pm }.$

\section{Generalized Einstein relation}

The diffusion coefficient of a normal random walk process can be related to
its mobility. This coefficient gives the proportionality between the force
and the average velocity of the walker when submitted to an external field.
This is the well known Einstein (fluctuation-dissipation) relation \cite%
{vanKampen,gardiner,bouchaud}. For the present model, it is not possible to
establishing a similar relation in terms of the ensemble behavior. In fact,
the different time dependences of the first two moments [see Eqs. (\ref%
{First}) and (\ref{third})] confirm this limitation. Given the ergodicity in
mean value defined by Eq. (\ref{MeanErgodic}) the same drawback applies to
the asymptotic time-averaged moments. Nevertheless, from the correlation
defined by Eq. (\ref{correlacionadas}) we realize that such kind of relation
can be obtained by introducing a centered (second) time-averaged moment
(second time-averaged cumulant), defined as%
\begin{equation}
\delta _{2}^{\ast }(t,\Delta )\equiv \delta _{2}(t,\Delta )-[\delta
_{1}(t,\Delta )]^{2}.  \label{centered}
\end{equation}%
Here, $\delta _{\kappa }(t,\Delta )$ $(\kappa =1,2)$ are the usual
time-averaged moments, Eq. (\ref{TimeAverage}). Denoting its asymptotic
value as%
\begin{equation}
\delta _{2}^{\ast }(\Delta )\equiv \lim_{t\rightarrow \infty }\delta
_{2}^{\ast }(t,\Delta ),
\end{equation}%
its average over an ensemble of realizations can be written as%
\begin{equation}
\left\langle \delta _{2}^{\ast }(\Delta )\right\rangle =\left\langle \delta
_{2}^{\ast }(\Delta ,f_{\pm })\right\rangle ,
\end{equation}%
where $\delta _{2}^{\ast }(\Delta ,f_{\pm }),$ from Eqs. (\ref{Delta1}) and (%
\ref{Delta2}), reads%
\begin{equation}
\delta _{2}^{\ast }(\Delta ,f_{\pm })=\delta
x^{2}[1-(f_{+}-f_{-})^{2}]\Delta .
\end{equation}%
In contrast to $\delta _{2}(\Delta ,f_{\pm })$ [Eq. (\ref{Delta2})], here a
linear dependence with $\Delta $ is obtained. Similarly, using that $%
\left\langle (f_{+}-f_{-})^{2}\right\rangle =[1+\lambda
(q_{+}-q_{-})^{2}]/(1+\lambda ),$ the average over realizations becomes%
\begin{equation}
\left\langle \delta _{2}^{\ast }(\Delta )\right\rangle =\delta x^{2}\frac{%
\lambda }{1+\lambda }[1-(q_{+}-q_{-})^{2}]\Delta .
\end{equation}

The case $q_{+}=q_{-}$ and $q_{+}\neq q_{-}$ define the unforced and forced
(driven) dynamics respectively. Taking a dimensional delay time $(\Delta
\rightarrow \Delta /\delta t),$ from the previous expression and Eq. (\ref%
{Delta1Media}) it follows%
\begin{equation}
\left\langle \delta _{2}^{\ast }(\Delta )\right\rangle
_{q_{+}=q_{-}}=2D_{\ast }\Delta ,\ \ \ \ \ \left\langle \delta _{1}(\Delta
)\right\rangle _{q_{+}\neq q_{-}}=V\Delta ,  \label{estrella}
\end{equation}%
where the (average) diffusion and (average) velocity coefficients are
[compare with Eq. (\ref{Continuo})]%
\begin{equation}
D_{\ast }\equiv \frac{1}{2}\frac{\delta x^{2}}{\delta t}\frac{\lambda }{%
1+\lambda },\ \ \ \ \ \ \ \ V\equiv \frac{\delta x}{\delta t}(q_{+}-q_{-}).
\end{equation}%
They can be related as%
\begin{equation}
D_{\ast }=\frac{\lambda }{1+\lambda }\frac{\delta x}{2}\frac{V}{(q_{+}-q_{-})%
},  \label{DEstrella}
\end{equation}%
which defines an Einstein-like relation. In fact, it relates the diffusion
coefficient corresponding to the centered (second) time-averaged moment of
the unforced dynamics with the velocity of the first time-averaged moment
for the forced case, Eq. (\ref{estrella}).

The standard Einstein relation involves a thermodynamic temperature \cite%
{vanKampen,gardiner,bouchaud}. Here, this dependence can be introduced by
assuming that the probabilities $q_{\pm }$ are given by a Boltzmann
exponential factor (activated process) $q_{\pm }=\exp [\pm \delta xF/2kT]/Z$ 
\cite{bouchaud}, where $F$ is the external force, $T$ the temperature, $k$
the Boltzmann constant, while $Z$ guarantee the normalization $%
q_{+}+q_{-}=1. $ Thus,%
\begin{equation}
q_{+}-q_{-}=\tanh \left[ \frac{\delta xF}{2kT}\right] .  \label{Hiperbola}
\end{equation}%
In the limit $F\rightarrow 0,$ Eqs. (\ref{DEstrella}) and (\ref{Hiperbola})
lead to%
\begin{equation}
D_{\ast }=\frac{\lambda }{1+\lambda }kT\left( \frac{V}{F}\right) .
\label{final}
\end{equation}%
In the limit $\lambda \rightarrow \infty ,$ it follows the standard Einstein
relation (see for example equation (5.3) in Ref. \cite{bouchaud}). In fact, $%
V/F$ is the (average) mobility. For finite $\lambda ,$ the standard result
is modified by the memory of the dynamics, which introduces the factor $%
\lambda /(1+\lambda ).$ Furthermore, notice that the generalized relation (%
\ref{final}) does not characterize the ensemble dynamics. In \ fact, it can
only be established in terms of the time-averaged moments [Eq. (\ref%
{estrella}], which satisfy%
\begin{equation}
\left\langle \delta _{1}(\Delta )\right\rangle _{q_{+}\neq q_{-}}=\frac{%
1+\lambda }{\lambda }\frac{(q_{+}-q_{-})}{\delta x}\left\langle \delta
_{2}^{\ast }(\Delta )\right\rangle _{q_{+}=q_{-}}.
\end{equation}%
From Eq. (\ref{Hiperbola}) this relation can be written as $(F\rightarrow 0)$%
\begin{equation}
\left\langle \delta _{1}(\Delta )\right\rangle _{F\neq 0}=\frac{1+\lambda }{%
\lambda }\frac{F}{2kT}\left\langle \delta _{2}^{\ast }(\Delta )\right\rangle
_{F=0}.
\end{equation}%
A similar property was also found for subdiffusive continuous-time random
walk models \cite{burovS} and others anomalous diffusion processes \cite%
{nogo,akimoto}.

\section{Summary and Conclusions}

We introduced a discrete random walk model driven by global memory effects,
where each walker step depends on the previous number of performed
left-rigth transitions, Eq. (\ref{Transition}). After obtaining a recursive
relation for its characteristic function, we obtained its firsts moments.
Given that the memory mechanism may induce a bias, the first moment has a
linear dependence with time, Eq. (\ref{First}). The second moment, event in
absence of bias, develops a superdiffusive ballistic behavior, Eq. (\ref%
{Second}). In a continuous time-space limit, the probability density is
governed by a (non-Markovian) local in-time Fokker-Planck equation [Eq. (\ref%
{FP})], being defined by an effective harmonic oscillator potential with a
strength constant inversely proportional to the elapsed time.

In the long time regime each realization is equivalent to that of a biased
Markovian walker with transitions rates that differs from realization to
realization. This kind of asymptotic inhomogeneous diffusion is induced by
the memory effects. Consequently, and similarly to the case of subdiffusive
continuous-time random walks, the time-averaged moments [Eq. (\ref%
{TimeAverage})] become random objects [Figs. (1) and (2)] with a time
independent statistics. Their average over realizations recover the ensemble
behavior obtained from the characteristic function [Fig. (3)]. Nevertheless,
due to their intrinsic randomness, characterized through their probability
densities [Figs. (4) and (5)], the diffusion process is nonergodic. For the
second-averaged moment we find that the ergodic limit and the limit of large
delay times do not commutate [Eq. (\ref{NoConmutan})]. Added to their
randomness, we showed that in general the time-averaged moments are
correlated between all them.

Due to the different time dependences of the first and second moments, it is
not possible to establish an Einstein-like relation for the ensemble
dynamics. Nevertheless, we showed that a generalized relation can be
formulated after introducing a centered (second) time-averaged moment
(second time-averaged cumulant), Eq. (\ref{centered}). In contrast with the
standard result, the relation between the corresponding (average) diffusion
and (average) mobility coefficients is modified by the memory control
parameter [Eqs. (\ref{DEstrella}) and (\ref{final})].

The present results, as well as the analyzes performed in Refs. \cite%
{deng,file,marcin,tejedor}, confirm that different kind of memory processes
may lead to weak ergodicity breaking, in particular that characterized by
random time-averaged moments (inhomogeneous diffusion). It is expected that
the same kind of results arise in continuous (time and space) random walk
models with finite residence times and finite average jump lengths. On the
other hand, conditions that guarantees that a memory mechanism leads (or
not) to ergodicity breaking are not known. General criteria for solving this
issue, as well as the interplay between global memory effects an divergent
residence times, jointly with the validity of the Einstein relation, are
interesting questions that emerge from the present analysis.

\section*{Acknowledgments}

This work was supported by Consejo Nacional de Investigaciones Cient\'{\i}%
ficas y T\'{e}cnicas (CONICET), Argentina.

\appendix

\section{\label{Beta}Probability density of the asymptotic transition
probabilities}

Here, we derive the probability density (\ref{Betal}) of the asymptotic
transition probabilities, $f_{\pm }=\lim_{t\rightarrow \infty }\mathcal{T}%
(\sigma _{1},\cdots \sigma _{t}|\sigma _{t+1}=\pm \delta x),$ Eq. (\ref%
{fractions}).

The joint probability $P(\sigma _{1},\cdots \sigma _{t})$ of obtaining the
random values $\sigma _{1},\cdots \sigma _{t},$\ by using Bayes rule, can be
written as%
\begin{equation}
P(\sigma _{1},\cdots \sigma _{t})\!=\!P(\sigma _{1})\mathcal{T}(\sigma
_{1}|\sigma _{2})\cdots \mathcal{T}(\sigma _{1},\cdots ,\sigma _{t-1}|\sigma
_{t}).
\end{equation}%
Given the transition probability Eq. (\ref{Transition}), it is simple to
check that $P(\sigma _{1},\cdots \sigma _{t})$ only depends on the number of
times $t_{\pm }$ that the values $\pm \delta x$ were chosen. From this
interchangeability property, the probability $P_{t}(t_{+},t_{-})$ of getting 
$t_{\pm }$ times the values $\pm \delta x$ after $t$ steps $(t=t_{+}+t_{-}),$
can be written as%
\begin{equation}
P_{t}(t_{+},t_{-})=\frac{t!}{t_{+}!t_{-}!}\frac{\Gamma (\lambda )}{\Gamma
(t+\lambda )}\frac{\Gamma (t_{+}+\lambda _{+})}{\Gamma (\lambda _{+})}\frac{%
\Gamma (t_{-}+\lambda _{-})}{\Gamma (\lambda _{-})},  \label{salio}
\end{equation}%
where the property $\Gamma (n+x)/\Gamma (x)=x(1+x)(2+x)\cdots (n-1+x)$ was
used. The combinatorial factor takes into account all realizations with the
same numbers $t_{\pm }.$

In the limit $x\rightarrow \infty $ it is valid the Stirling approximation $%
\Gamma (x)\approx \sqrt{2\pi /x}e^{-x}x^{x},$ which in the same limit leads
to $\Gamma (x+\alpha )/\Gamma (x)\approx x^{\alpha }.$ Using that $n!=\Gamma
(n+1),$ and applying the previous approximations to Eq. (\ref{salio}), in
the limit $t\rightarrow \infty $ it follows%
\begin{equation}
P_{t}(t_{+},t_{-})\approx \frac{\Gamma (\lambda )}{t^{\lambda -1}}\frac{%
t_{+}^{\lambda _{+}-1}}{\Gamma (\lambda _{+})}\frac{t_{-}^{\lambda _{-}-1}}{%
\Gamma (\lambda _{-})}.
\end{equation}%
By performing the change of variables $t_{\pm }\rightarrow tf_{\pm },$ and
by using that, due to normalization $t=t_{+}+t_{-},$ there is only one
independent variable $(f_{+}+f_{-}=1),$ the previous expression
straightforwardly leads to the Beta distribution Eq. (\ref{Betal}).

\section{Ergodicity in mean value}

Here, we demonstrate the validity of Eqs. (\ref{MeanErgodic}) and (\ref%
{MeanErgodicBis}). Their fulfilment imply that the random walk is ergodic in
mean value. The demonstration has a close relation with the de Finetti
representation theorem for dichotomic variables \cite{feller,budini}. In the
present context, we notice that the probability $P_{t}(t_{+},t_{-})$ [Eq. (%
\ref{salio})] can be written as%
\begin{equation}
P_{t}(t_{+},t_{-})=\int_{0}^{1}df_{+}\ \mathcal{P}(f_{\pm
})P_{t}(t_{+},t_{-},f_{\pm }).  \label{PnAsAverage}
\end{equation}%
Here, $\mathcal{P}(f_{\pm })$ is given by Eq. (\ref{Betal}) while $%
P_{t}(t_{+},t_{-},f_{\pm })$ is the counting probability for independent
variables $\sigma _{i}=\pm \delta x$ with transition probability $\mathcal{T}%
(\sigma _{1},\cdots \sigma _{t}|\sigma _{t+1}=\pm \delta x)=f_{\pm }.$
Therefore, it is%
\begin{equation}
P_{t}(t_{+},t_{-},f_{\pm })\equiv \frac{t!}{t_{+}!t_{-}!}%
f_{+}^{t_{+}}f_{-}^{t_{-}}.
\end{equation}%
Given that the characteristic function $Q_{t}(k)$ [Eq. (\ref{Qk})] can be
written as%
\begin{equation}
Q_{t}(k)=\sum_{t_{\pm }=0}^{t}P_{t}(t_{+},t_{-})\exp [ik\delta
x(t_{+}-t_{-})],
\end{equation}%
where $t_{+}+t_{-}=t,$ Eq. (\ref{PnAsAverage}) allows us to write $Q_{t}(k)$
as an average over the variables $f_{\pm }$%
\begin{equation}
Q_{t}(k)=\int_{0}^{1}df_{+}\ \mathcal{P}(f_{\pm })Q_{t}(k,f_{\pm }),
\label{QbyFinetti}
\end{equation}%
where $Q_{t}(k,f_{\pm })$ is the characteristic function for independent
variables with transition probabilities $f_{\pm },$%
\begin{equation}
Q_{t}(k,f_{\pm })=[f_{+}e^{+ik\delta x}+f_{-}e^{-ik\delta x}]^{t}.
\label{QIndependent}
\end{equation}%
Given that \textit{asymptotically} the realizations of the random walk
converge to that of a memoryless process with transition rate $\mathcal{T}%
(\sigma _{1},\cdots \sigma _{t}|\sigma _{t+1})=f_{\pm }$ [Eq. (\ref%
{fractions})], in each realization the (asymptotic) statistics of $%
[x(t^{\prime }+\Delta )-x(t^{\prime })],$ which define the integral defining 
$\delta _{\kappa }(t,\Delta )$ [Eq. (\ref{TimeAverage})], does not depends
on $t$ and is defined by Eq. (\ref{QIndependent}) under the replacement $%
t\rightarrow \Delta .$ The relation Eq. (\ref{MeanErgodicBis}) is a
straightforward consequence on this result and Eq. (\ref{QbyFinetti}).


\begin{thebibliography}{99}
\bibitem{previol} The sum contributions in Eq. (\ref{previa}) can be
rewritten as $\sum_{\mu =\pm }q_{\mu }e^{ik\delta x_{\mu }}=\cos (k\delta
x)+i(q_{+}-q_{-})\sin (k\delta x),$ and as $\sum_{\mu =\pm }\left\langle
e^{ikx_{t}}t_{\mu }\right\rangle e^{ik\delta x_{\mu }}=tQ_{t}(k)\cos
(k\delta x)+i\left\langle e^{ikx_{t}}(t_{+}-t_{-})\right\rangle \sin
(k\delta x).$
\bibitem{fourier} We used that, given the Fourier transform $\tilde{f}%
(k)=\cos (ka)f(k),$ then $\tilde{f}(x)=[f(x-a)+f(x+a)]/2.$ When $\tilde{f}%
(k)=i\sin (ka)f(k),$ it follows $\tilde{f}(x)=[f(x-a)-f(x+a)]/2.$ If $\tilde{%
f}(k)=\frac{\sin (ka)}{a}\frac{df(k)}{dk},$ then $\tilde{f}(x)=\frac{(x-a)}{%
2a}f(x-a)-\frac{(x+a)}{2a}f(x+a)].$
\end{thebibliography}
\end{document}